\documentclass[12pt]{article}
\usepackage[latin1]{inputenc}
\usepackage{amsfonts}
\usepackage{amssymb}
\usepackage{graphics,psboxit,amsmath, epsfig}
\usepackage{psfrag}
\usepackage{verbatim}
\usepackage{fancyhdr}

\def\hybrid{\topmargin -20pt    \oddsidemargin 0pt
        \headheight 0pt \headsep 0pt
        \textwidth 6.35in       
        \textheight 9.25in       
        \marginparwidth .875in
        \parskip 5pt plus 1pt   \jot = 1.5ex}

\hybrid

\def\baselinestretch{1.2}

\catcode`\@=11
\def\marginnote#1{}
\newcount\hour
\newcount\minute
\newtoks\amorpm
\hour=\time\divide\hour by60
\minute=\time{\multiply\hour by60 \global\advance\minute by-\hour}
\edef\standardtime{{\ifnum\hour<12 \global\amorpm={am}%
        \else\global\amorpm={pm}\advance\hour by-12 \fi
        \ifnum\hour=0 \hour=12 \fi
        \number\hour:\ifnum\minute<10 0\fi\number\minute\the\amorpm}}
\edef\militarytime{\number\hour:\ifnum\minute<10 0\fi\number\minute}

\def\draftlabel#1{{\@bsphack\if@filesw {\let\thepage\relax
   \xdef\@gtempa{\write\@auxout{\string
      \newlabel{#1}{{\@currentlabel}{\thepage}}}}}\@gtempa
   \if@nobreak \ifvmode\nobreak\fi\fi\fi\@esphack}
        \gdef\@eqnlabel{#1}}
\def\@eqnlabel{}
\def\@vacuum{}
\def\draftmarginnote#1{\marginpar{\raggedright\scriptsize\tt#1}}

\def\draft{\oddsidemargin -.5truein
        \def\@oddfoot{\sl preliminary draft \hfil
        \rm\thepage\hfil\sl\today\quad\militarytime}
        \let\@evenfoot\@oddfoot \overfullrule 3pt
        \let\label=\draftlabel
        \let\marginnote=\draftmarginnote
   \def\@eqnnum{(\theequation)\rlap{\kern\marginparsep\tt\@eqnlabel}%
\global\let\@eqnlabel\@vacuum}  }

\def\preprint{\twocolumn\sloppy\flushbottom\parindent 2em
        \leftmargini 2em\leftmarginv .5em\leftmarginvi .5em
        \oddsidemargin -.5in    \evensidemargin -.5in
        \columnsep .4in \footheight 0pt
        \textwidth 10.in        \topmargin  -.4in
        \headheight 12pt \topskip .4in
        \textheight 6.9in \footskip 0pt
        \def\@oddhead{\thepage\hfil\addtocounter{page}{1}\thepage}
        \let\@evenhead\@oddhead \def\@oddfoot{} \def\@evenfoot{} }

\def\numberbysection{\@addtoreset{equation}{section}
        \def\theequation{\thesection.\arabic{equation}}}

\def\underline#1{\relax\ifmmode\@@underline#1\else
        $\@@underline{\hbox{#1}}$\relax\fi}

\def\titlepage{\@restonecolfalse\if@twocolumn\@restonecoltrue\onecolumn
     \else \newpage \fi \thispagestyle{empty}\c@page\z@
        \def\thefootnote{\fnsymbol{footnote}} }

\def\endtitlepage{\if@restonecol\twocolumn \else \newpage \fi
        \def\thefootnote{\arabic{footnote}}
        \setcounter{footnote}{0}}  

\catcode`@=12
\relax

\def\figcap{\section*{Figure Captions\markboth
        {FIGURECAPTIONS}{FIGURECAPTIONS}}\list
        {Figure \arabic{enumi}:\hfill}{\settowidth\labelwidth{Figure
999:}
        \leftmargin\labelwidth
        \advance\leftmargin\labelsep\usecounter{enumi}}}
 \relax
\def\tablecap{\section*{Table Captions\markboth
        {TABLECAPTIONS}{TABLECAPTIONS}}\list
        {Table \arabic{enumi}:\hfill}{\settowidth\labelwidth{Table
999:}
        \leftmargin\labelwidth
        \advance\leftmargin\labelsep\usecounter{enumi}}}
 \relax
\def\reflist{\section*{References\markboth
        {REFLIST}{REFLIST}}\list
        {[\arabic{enumi}]\hfill}{\settowidth\labelwidth{[999]}
        \leftmargin\labelwidth
        \advance\leftmargin\labelsep\usecounter{enumi}}}
 \relax
\makeatletter
\newcounter{pubctr}
\def\publist{\@ifnextchar[{\@publist}{\@@publist}}
\def\@publist[#1]{\list
        {[\arabic{pubctr}]\hfill}{\settowidth\labelwidth{[999]}
        \leftmargin\labelwidth
        \advance\leftmargin\labelsep
        \@nmbrlisttrue\def\@listctr{pubctr}
        \setcounter{pubctr}{#1}\addtocounter{pubctr}{-1}}}
\def\@@publist{\list
        {[\arabic{pubctr}]\hfill}{\settowidth\labelwidth{[999]}
        \leftmargin\labelwidth
        \advance\leftmargin\labelsep
        \@nmbrlisttrue\def\@listctr{pubctr}}}
 \relax
\makeatother
\newskip\humongous \humongous=0pt plus 1000pt minus 1000pt

\newif\ifdtup

\relax

\def\be{\begin{equation}}
\def\ee{\end{equation}}
\def\ba{\begin{eqnarray}}
\def\ea{\end{eqnarray}}

\def\no{\noindent}

\def\IR{\relax{\rm I\kern-.18em R}}
\def\II{\relax{\rm 1\kern-.35em1}}

\renewcommand{\theequation}{\thesection.\arabic{equation}}
\csname @addtoreset\endcsname{equation}{section}

\def\N4{${\cal N}=4$}

\def\bo{\boldsymbol }
\newcommand{\PD}[1]{\frac{\partial}{\partial #1}}

\newcommand{\Feyn}[1]{#1\kern-0.45em/}

\begin{document}

\title{\Large \bf Commuting Conformal and Dual Conformal Symmetries in the Regge limit}
\author{\large Johan~Gunnesson \\
{\it Instituto de F{\' i}sica Te{\' o}rica UAM/CSIC,}\\ 
{\it Universidad Aut{\' o}noma de Madrid, E-28049 Madrid, Spain}}

\maketitle

\vspace{-9cm}
\begin{flushright}
{\small IFT--UAM/CSIC--10-12}
\end{flushright}

\vspace{7cm}
\begin{abstract}
\noindent


In this paper we continue our study of the dual SL(2,C) symmetry of the BFKL equation, analogous to the dual conformal symmetry of N=4 Super Yang Mills. We find that the ordinary and dual SL(2,C) symmetries do not generate a Yangian, in contrast to the ordinary and dual conformal symmetries in the four-dimensional gauge theory. The algebraic structure is still reminiscent of that of N=4 SYM, however, and one can extract a generator from the dual SL(2,C) close to the bi-local form associated with Yangian algebras. We also discuss the issue of whether the dual SL(2,C) symmetry, which in its original form is broken by IR effects, is broken in a controlled way, similar to the way the dual conformal symmetry of N=4 satisfies an anomalous Ward identity. At least for the lowest orders it seems possible to recover the dual SL(2,C) by deforming its representation, keeping open the possibility that it is an exact symmetry of BFKL.
   
\end{abstract}

\section{Introduction}

\no
Almost two decades ago it was shown that the generalized Leading Logarithmic Approximation in the Regge limit of QCD can be described by a spin chain \cite{GLLAspinchain}, which was later shown in \cite{FaddeevKorchemsky} to be completely integrable, i.e. have an infinite tower of conserved charges allowing an analytic diagonalization of its Hamiltonian. The spin chain is defined by a nearest-neighbor interaction given by the BFKL hamiltonian, which has been known since the 70's as the kernel of the BFKL equation \cite{BFKL}, giving the Leading Logarithmic Approximation to scattering with color singlet exchange.

From the viewpoint of QCD it is surprising that such an infinite amount of symmetry is found in the Regge limit. The simplest explanation probably lies in that the Regge limit of QCD coincides to LLA with the Regge limit of the maximally supersymmetric extension of pure Yang-Mills, $\mathcal{N}=4$ Super Yang Mills, for which there is a large amount of evidence that it is integrable (in the large $N_c$ limit) to all orders in the gauge coupling \cite{integrabilityN4}. Having an exact solution of $\mathcal{N}=4$ SYM will induce a solution in the Regge limit, which would explain its integrability. To date, however, there is little understanding of how the integrable structure of $\mathcal{N}=4$ implies the symmetries of BFKL and its extensions.

Part of the difficulty in making the connection lies in that the symmetries in the Regge limit are naturally understood in terms of  a two dimensional effective theory (see for example \cite{ReggeAction}), in which the longitudinal components of the momenta have decoupled and the dynamics of the theory have been reduced to the transverse plane. An essential component of the symmetry found in this limit is given by the two-dimensional conformal $SL(2,C)$ symmetry discovered by Lipatov \cite{Lipatov1986}. The running of the coupling constant does not enter the LLA, and one can thus suspect that this symmetry is a consequence of the classical conformal symmetry of the QCD Lagrangian. When effects from the running of the coupling are included at NLLA, the symmetry is broken. In $\mathcal{N}=4$ SYM, by contrast, the coupling does not run, and recent work indeed seems to indicate that the $SL(2,C)$ symmetry remains exact at NLLA \cite{BFKLN4}. It therefore seems natural to understand the $SL(2,C)$ as what remains of the four-dimensional conformal symmetry of $\mathcal{N}=4$ when taking the Regge limit. 

The infinite Yangian algebra \cite{Yangian} of conserved charges of $\mathcal{N}=4$ is generated by two copies of the superconformal group $PSU(2,2|4)$ \cite{beisertyangian, plefka}, the first being the ordinary superconformal symmetry of the Lagrangian, while the second is a novel, non-Lagrangian symmetry, coined dual superconformal symmetry \cite{dualconformal, confwardident}. If it were possible, in the same way that the ordinary conformal symmetry would seem to imply the ordinary $SL(2,C)$, to identify a non-trivial remnant of the dual conformal symmetry in the Regge limit, one could quite possibly explain the integrability of the generalized LLA as being generated by the ordinary $SL(2,C)$ and this new dual symmetry \footnote{The dual conformal symmetry is also probably related to the symmetry, similar to the dual $SL(2,C)$ of this article, used in \cite{Lipatovopenintegrable} to find the spectrum of an integrable, open spin chain in the octet channel.}.

In \cite{Nosotros} we took a step in this direction, identifying a dual $SL(2,C)$ symmetry of BFKL. However, as discussed in that article, the most straightforward application of the dual $SL(2,C)$ turns out to be broken by IR effects, which produce anomalous terms when iterating the BFKL equation. In this article, we discuss whether it is possible that such terms can be reabsorbed into a deformation of the symmetry. This would parallel the way that the dual conformal symmetry of $\mathcal{N}=4$ is made exact by correcting for IR anomalies \cite{confwardident}. We also initiate a study of the algebra that is obtained by commuting the ordinary and dual $SL(2,C)$ symmetries, extracting a piece of the dual symmetry which is close to a bi-local form frequently appearing in Yangian algebras. Although not entirely surprising, the Serre relations, which define a Yangian algebra, are however not satisfied.

\section{The dual $SL(2,C)$ symmetry}

\begin{figure}[ht]

\psfrag{p1}{$p_1$} \psfrag{p2}{$p_2$} \psfrag{p3}{$p_3$} \psfrag{p4}{$p_4$}
\psfrag{k}{$k$} \psfrag{kpp4}{$k+p_4$} \psfrag{kmp1}{$k - p_1$} \psfrag{kmp1mp2}{$k-p_1-p_2$}
\begin{center}
\includegraphics{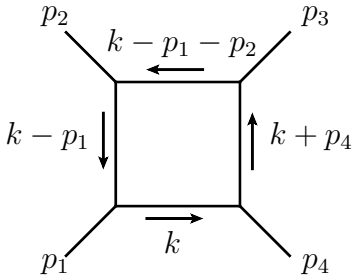}

\caption{\small The scalar box diagram, giving the one-loop correction to four particle scattering amplitude in $\mathcal{N}=4$ Super Yang Mills, in terms of incoming momenta $p_i$, and an integral over the loop momentum $k$.} \label{fig:scalarbox}
\end{center}
\end{figure}

\no
In this section we will motivate the definition of the dual $SL(2,C)$ symmetry. Since we are looking for what is left in the Regge limit of the dual conformal symmetry of $\mathcal{N}=4$ scattering amplitudes, let us first recall how this four-dimensional symmetry is constructed. If one introduces a set of variables 
$x_i$, related to the external (all taken as incoming) particle momenta $p_i$, $i=1\ldots n$, through
\be
p_i=x_i - x_{i+1}\equiv x_{i\, i+1} \ , \label{eq:xi}
\ee
the amplitudes exhibit, at tree level, covariance under a superconformal group acting on the $x$ variables in the same way as the ordinary superconformal symmetry acts on spatial coordinates \cite{dualconformal}. Loop integrals are then found to be formally (before IR regularization) invariant under the same symmetry. As an example, which will help motivate our version of the symmetry in the Regge limit, let us look at the one-loop correction to the four-particle amplitude. In $\mathcal{N}=4$ SYM it is simply proportional to the scalar box diagram \cite{scalarbox}, shown in figure \ref{fig:scalarbox}.

The integral which one obtains when evaluating this diagram is
\be
\int \frac{d^4k \, (p_1 + p_2)^2 (p_3+p_4)^2}{k^2 (k-p_1)^2(k-p_1-p_2)^2(k+p_4)^2} \ ,
\ee
which becomes, after introducing $x$-variables \eqref{eq:xi}, and defining a new integration variable $x_I$ through $k=x_{1I}$,
\be
\int \frac{d^4x_I \, x_{13}^2x_{24}^2}{x_{1I}^2x_{2I}^2x_{3I}^2x_{4I}^2} \ . \label{eq:xint}
\ee
If the external particles are on-shell, with $p_i^2=0$, this integral diverges, but ignoring this for the moment one has a formal conformal symmetry acting on the $x$. This is easily seen by noting that under conformal inversions $x_i^\mu \rightarrow \frac{x_i^\mu}{x_i^2}$ implying that the squared differences $\bo{x}_{ij}^2$, and the measure $d^4x_I$ transform as 
\be
x ^2_{ij} \rightarrow \frac{x_{ij}^2}{x_i^2x_j^2} \ , \;\; \text{and} \;\; d^4x_I\rightarrow \frac{d^4x_I}{x_I^8} \ , \label{eq:diffandmeasure}
\ee
respectively.

Let us now make contact with the Regge limit. In non-abelian gauge theories the gluon reggeizes, meaning that when the quantum numbers of the external states so permit, the amplitudes can be described in terms of effective particles propagating in the $t$-channel called reggeized gluons, which carry the quantum numbers of ordinary gluons, but for which the propagators (in Feynman gauge) have been replaced as 
\be
\frac{g_{\mu \nu}}{t}\rightarrow \frac{g_{\mu \nu}}{t}\left(\frac{s}{\bo{k}^2} \right)^{\omega (t)} \ , \label{eq:reggeprop}
\ee
where $\omega (t)$ is called the gluon Regge trajectory, and the scale $\bo{k}^2$ is of order $t$ but whose value does otherwise not matter at LLA. To leading order, the gluon trajectory reads
\be
\omega (\bo{q}^2) = -\frac{\bar{\alpha}}{4 \pi }\int d^2\bo{k}' \frac{\bo{q}^2}{\bo{k}'^2(\bo{k}'-\bo{q})^2} \ , 
\label{eq:trajectoryreggegluon}
\ee
where $\bar{\alpha}=\frac{g^2N_c}{4 \pi ^2 }$, with $g^2$ the gauge theory coupling and $N_c$ the number of colors. The form of the trajectory reflects, as mentioned in the introduction, that the dynamics are reduced to the two-dimensional transverse plane in the Regge limit, so that $t=-\bo{q}^2$, where $\bo{q}$ is the transverse part of the momentum exchanged in the scattering process. The substitution \eqref{eq:reggeprop} incorporates an all-loop resummation, but knowing that the gluon reggeizes it is enough to perform a one-loop computation to calculate the trajectory. Expanding \eqref{eq:reggeprop} in the coupling, it is clear that the one-loop amplitude for gluon quantum number exchange should in the Regge limit be given by the tree amplitude multiplied by $\omega (t)\ln \frac{s}{\bo{k}^2}$. Perhaps the simplest way to perform this calculation and obtain \eqref{eq:trajectoryreggegluon}, as reviewed nicely in \cite{rossforshaw}, is to use unitarity, the Cutkosky rules and dispersion relations, to obtain the 1-loop amplitude as a product of two tree amplitudes. This corresponds to setting the upper and lower propagators of figure \ref{fig:scalarbox} on-shell. Doing so removes the propagators $x_{1I}^2$ and $x_{3I}^2$ from the denominator of \eqref{eq:xint} and introduces two delta functions, thereby reducing the dimensions of the integral from four to two. Furthermore, in the Regge limit, the integral is dominated by the transverse components of the momenta, replacing the $x$ by two-dimensional vectors $\bo{x}$. Taking into consideration the tree-level factors (which corrects numerator of \eqref{eq:xint}) one then obtains the reggeized gluon trajectory as
\be
\omega (\bo{x}^2_{24})=-\frac{\bar{\alpha}}{4 \pi }\int d^2 \bo{x}_I \frac{\bo{x}^2_{24}}{\bo{x}^2_{2I}\bo{x}^2_{4I}} \ . \label{eq:trajectoryreggegluonk1x}
\ee

The same representation of the trajectory would have been obtained by shifting the integration variable $\bo{k}'$ of \eqref{eq:trajectoryreggegluon} as $\bo{x}_{2I}$, where $\bo{x}_I$ is again taken as the new integration variable, and performing the replacement  $\bo{q}=\bo{p}_2+\bo{p}_3\to \bo{x}_{24}$. This last replacement is precisely what one would obtain from a two-dimensional version of \eqref{eq:xi}. Furthermore, the integral representation \eqref{eq:trajectoryreggegluonk1x} has a formal (ignoring IR divergences) two-dimensional inversion symmetry
\be
\bo{x}_i\rightarrow \frac{\bo{x}_i}{\bo{x}^2} \ \label{eq:xinversion}
\ee
since the squared differences $x^2_{i  j}$ and the measure transforms, similarly to \eqref{eq:diffandmeasure}, as
\be
\bo{x} ^2_{ij} \rightarrow \frac{\bo{x}_{ij}^2}{\bo{x}_i^2\bo{x}_j^2} \ , \;\; \text{and} \;\; d^2\bo{x}_I \rightarrow \frac{d^2\bo{x}_I}{\bo{x}_I^4} \ . \label{eq:diffandmeasure2d}
\ee

This suggest that a remnant of the dual conformal symmetry can be found in the effective two-dimensional theory that arises in the Regge limit by simply introducing $\bo{x}$-variables, related to the incoming, transverse momenta $\bo{p}_i$ by
\be
\bo{p}_i=\bo{x}_i -\bo{x}_{i+1} \ , \label{eq:xi2d}
\ee
and then acting on the $\bo{x}$-variables in the standard way. Together with the inversions, one also has translation symmetry (since the $\bo{x}$ only enter as the differences \eqref{eq:xi2d}) as well as rotation symmetry. In total, this constitutes the group $SL(2,C)$.


Since our main motivation for looking at the dual $SL(2,C)$ was to try to understand the integrability of the generalized LLA spin chain, we should see what form it takes for color singlet exchange. As a first step one is led to search for a dual $SL(2,C)$ symmetry of the BFKL equation. This equation can be interpreted,  in the effective two-dimensional theory that arises in the Regge limit, as giving the leading approximation to the scattering amplitude $f$ of 2 to 2 reggeized gluons, with color singlet exchange in the $t$-channel. If we label the incoming momenta of these reggeized gluons as $\bo{p}_1 ,\, \ldots ,\, \bo{p}_4$, we can introduce $\bo{x}$-variables using \eqref{eq:xi2d}, in terms of which the BFKL equation takes the form
\begin{equation}
\omega f(\omega , \, \bo{x}_1,\, \ldots,\, \bo{x}_4) = \delta^{(2)}(\bo{x}_{24}) + \int d^2\bo{x}_I K(\bo{x}_{1},\, \bo{x}_{2},\, \bo{x}_{3},\, \bo{x}_{I}) f(\omega , \,  \bo{x}_1,\, \bo{x}_I,\, \bo{x}_3,\, \bo{x}_4) \, \label{eq:BFKL} 
\end{equation}
with the integration kernel given by
\be
K(\bo{x}_{1},\, \bo{x}_{2},\, \bo{x}_{3},\, \bo{x}_{I}) = 
\frac{\bar{\alpha}}{2\pi}\frac{K_R (\bo{x}_{12},\, \bo{x}_{23}; \bo{x}_{3I}, \, \bo{x}_{I1})}{ \bo{x}_{12}^2\bo{x}_{I3}^2} 
+ \left[  \omega (\bo{x}_{12}^2) + \omega (\bo{x}_{23}^2) \right]\delta ^{(2)}(\bo{x}_{I4}) \ , \label{eq:Kernelx}
\ee
where
\be
K_R (\bo{x}_{12},\, \bo{x}_{23}; \bo{x}_{3I}, \, \bo{x}_{I1}) = -\left[  \bo{x}_{13}^2 - \frac{\bo{x}_{23}^2 \bo{x}_{I1}^2}{\bo{x}_{2I}^2}- \frac{\bo{x}_{12}^2\bo{x}_{I3}^2}{\bo{x}_{2I}^2}  \right]  \  \label{eq:KernelRx}
\ee
and $\omega (x)$ is the gluon Regge trajectory. The variable $\omega$ is, despite the notation, which we keep for historical reasons, not the gluon trajectory, but rather the Mellin transform conjugate of $\frac{s}{\bo{k}^2}$. 


In \cite{Nosotros} we showed that \eqref{eq:BFKL} is invariant under a formal dual $SL(2,C)$ symmetry, which acts on the $x$ variables in the same way that the $SL(2,C)$ of Lipatov acts on impact parameter space. The symmetry requires that $\omega$ be invariant, which either means that we should not transform the longitudinal components of the momenta at all, or should require that the scale $\bo{k}^2$ transform in the same way as $s$.

\section{The IR anomaly}

\no
The dual $SL(2,C)$ is only a formal symmetry of the BFKL equation and the gluon Regge trajectory, and is expected to be affected by the IR divergences present. In BFKL the divergences cancel, so one could hope that the symmetry remains unaltered, but as mentioned in \cite{Nosotros} this is unfortunately not the case. The gluon Regge trajectory is given, in dimensional regularization, by
\be
\omega (\bo{x}^2_{12})=-\frac{\bar{\alpha}}{4 \pi }(4 \pi \mu)^{2\epsilon}\int d^{2-2\epsilon} \bo{x}_I \frac{\bo{x}^2_{12}}{\bo{x}^2_{I1} \bo{x}^2_{I2}}\approx -\frac{\bar{\alpha}}{2 }(4\pi e^{-\gamma})^\epsilon  \left( \log \frac{\bo{x}^2_{12}}{\mu ^2}- \frac{1}{\epsilon} \right) \ . \label{eq:trajectorydimreg}
\ee
In the BFKL equation the poles in $\epsilon$ and the logarithms in the scale $\mu^2$ cancel, but factors such as $\log \bo{x}^2_{12}$ do not cancel and rather add up to an anomalous, non-invariant expression.

Perhaps the most direct way to see this is to write out the first orders to the solution $f$ of the equation, since a consequence of the symmetry, if it where to remain exact, is that
\be
f \rightarrow \bo{x}^2_2\bo{x}_4^2 f \label{eq:Ftransform}
\ee
under dual inversions \eqref{eq:xinversion}.

The solution to BFKL can be constructed, order by order, by iterating the equation, regularizing and canceling infrared divergences at each order. In general, the solution can be written
\be
\omega f = f_1(\hat{\alpha}) \, \delta ^{(2)}(\bo{x}_{24})+\frac{1}{\bo{x}^2_{24}}f_2(\hat{\alpha}) \ ,
\ee
where $\hat{\alpha}\equiv \bar{\alpha}/\omega$, $f_1 (\hat{\alpha}) $ is obtained by repeatedly iterating the trajectory part of the kernel, and where $f_2(\hat{\alpha})$ must be invariant under dual inversions in order for \eqref{eq:Ftransform} to hold. The lowest order of $F_2$ is trivial to calculate, simply being the result of applying the kernel \eqref{eq:Kernelx} to the inhomogenous delta-function term of \eqref{eq:BFKL}. At the next order, apart from integrals similar to the trajectory, non-trivial integrals of the form
\be
\frac{1}{\pi}\int \frac{d^2\bo{x}_I \, \bo{x}_{1I}^2}{\bo{x}_{2I}^2 \bo{x}_{3I}^2 \bo{x}_{4I}^2} \ 
\ee
appear, but as shown in appendix \ref{app:nfintegral} their form can be considerably restricted by studying their symmetry properties under dual $SL(2,C)$ transformations.  One finds that
\be
f_2(\hat{\alpha})  = \frac{\hat{\alpha}}{2 \pi}\left( 1 + u -v \right)+\frac{\hat{\alpha}^2}{2 \pi}\left[\left( 1 + u -v \right)  \log \left( \frac{\bo{x}_{24}^4}{\bo{x}_{12}^2 \bo{x}_{34}^2} \right) +v \log v - u \log u \right] + \cdots \ , \label{eq:F2}
\ee
where
\be
u = \frac{\bo{x}_{14}^2\bo{x}_{23}^2}{\bo{x}_{12}^2\bo{x}_{34}^2} \; \;\;\ \text{and} \;\;\; v = \frac{\bo{x}_{13}^2\bo{x}_{24}^2}{\bo{x}_{12}^2\bo{x}_{34}^2}
\ee
are the two independent conformal invariants that can be formed from $\bo{x}_1, \, \ldots ,\, \bo{x}_4$. We see that the lowest order is invariant, while the next-to-lowest order, where the trajectories start to matter, is not, containing an anomalous logarithmic term \footnote{The symmetry is still present for the special kinematical configurations satisfying $1+u=v$, but it is simple to check numerically that even for such configurations $f_2$ is not invariant at higher orders.}.

Such a logarithm is not completely unexpected, though, since the dual conformal symmetry of  $\mathcal{N}=4$ SYM is broken by infrared divergences. In that case it seems, fortunately, to be broken in a controlled way. An $n$-particle color-ordered $\mathcal{N}=4$ scattering amplitude $\mathcal{A}_n$ can always be factorized as
\be
\mathcal{A}_n = \mathcal{A}_{n,\text{tree}} \mathcal{M}_{n} \ ,
\ee
where $\mathcal{A}_{n,\text{tree}}$ is the tree-level amplitude. 
There is evidence that the finite part of $\mathcal{M}_{n} $ satisfies to all orders an "anomalous Ward identity" \cite{confwardident} given by
\be
K^{\mu} \log \mathcal{M}_{n,\text{finite}}=\frac{1}{2}\Gamma _{\text{cusp}}\left(\frac{\bar{\alpha}}{2} \right)\sum _{i=1}^n \log \frac{x_{i,i+2}^2}{x^2_{i-1,i+1}}x_{i,i+1}^\mu \ , \label{eq:ward}
\ee 
where $\Gamma _{\text{cusp}}\left(\frac{\bar{\alpha}}{2} \right)$ is the cusp anomalous dimension.

If we instead apply the generator $K_{\mu}^{(D)}$ (to be defined in the next section) of the dual $SL(2,C)$ corresponding to the special conformal transformations  to $f_2$ we obtain (to order $\hat{\alpha}$)
\be
i K_\mu^{(D)} \log f_2 = 2\gamma_1(\hat{\alpha}) \left(x_{1\mu} + x_{3\mu} - x_{2\mu} - x_{4\mu} \right) \ , \label{eq:ward2d}
\ee
 where $\mu$ here is a two-dimensional index, where $\gamma_1(\hat{\alpha}) = \hat{\alpha} + \mathcal{O}\left(  \hat{\alpha}^2 \right)$, and where we omit the bold font from the $\bo{x}$ when writing them out in components. This is a rather simple structure, which stems from the fact that the anomalous term is a simple logarithm, to lowest order, while the anomalous terms in the finite part of $\mathcal{N}=4$ amplitudes consist of squares of logarithms. The important question is now if this structure is exact to all loops, as seems to be the case in the four-dimensional gauge theory.
 
Unfortunately, it is not. Perhaps the simplest way to see this analytically is to consider the forward limit, defined by $\bo{q}=0$, or equivalently $\bo{x}_1=\bo{x}_3$, where calculations simplify considerably. It is then not difficult to calculate the third order of $f_2$. In the forward limit, $u=1$ and $v=0$, and \eqref{eq:ward2d} becomes
\be
i K_\mu^{(D)} \log f_{2,\text{forward}} = 2 \gamma_1 (\hat{\alpha}) ( 2 x_1^\mu  - x_2^\mu - x_4^\mu)  \ , \label{eq:ward2dforward}
\ee
For this to hold to all orders, in the forward limit $\log f_2$ would have to be of the form
\be
\gamma _1 (\hat{\alpha})  \log \left( \frac{\bo{x}_{24}^4}{\bo{x}_{12}^2 \bo{x}_{14}^2} \right) + \gamma _2(\hat{\alpha}) \ ,
\ee
where $\gamma _1$ and $\gamma_2$ only depend on $\hat{\alpha}$. Instead, one obtains
\be
f_{2,\text{forward}} (\hat{\alpha})  = \frac{\hat{\alpha}}{\pi}+\frac{\hat{\alpha}^2}{\pi} \log \left( \frac{\bo{x}_{24}^4}{\bo{x}_{12}^2 \bo{x}_{14}^2} \right) + \frac{\hat{\alpha}^3}{\pi} \left[ \log ^2 \left( \frac{\bo{x}_{24}^4}{\bo{x}_{12}^2 \bo{x}_{14}^2} \right) - \frac{1}{4}   \log ^2 \frac{\bo{x}_{12}^2}{\bo{x}_{14}^2} - \frac{\pi^2}{3}  \right]  +  \cdots \ . \label{eq:f2forward}
\ee
As a result \eqref{eq:ward2d} does not hold to all orders in BFKL. This does not mean that the dual $SL(2,C)$ symmetry is broken beyond repair, however. It simply means that the way that the representation of the symmetry is deformed by the coupling is not as simple as in the case of $\mathcal{N}=4$. 

One way to understand that the structure of the representation of the dual $SL(2,C)$ may be complicated even though it may stem from the simple dual conformal symmetry is to visualize, in the standard way, the iteration of the BFKL equation as a sum over effective ladder diagrams. These diagrams can then, in turn, be written, by unitarity, as a product of tree-level amplitudes. Summing over the rungs of the ladder will then sum over products of tree-amplitudes with a different number of legs, and consequently different dual conformal weights. 

An alternative way to see this is to regularize the BFKL equation in such a way that the dual $SL(2,C)$ invariance is preserved by adapting the procedure of \cite{AldayHiggs}, introducing different scales which are allowed to transform independently under inversions. Then one finds that new scales must be introduced when iterating the equation. Incidentally, this means that without a more detailed description of how the dual $SL(2,C)$ is represented at higher orders it cannot be used to constrain the form of $f$.

Still, we believe that it is possible to deform the representation of the dual $SL(2,C)$ in such a way to make it an all order symmetry.   At the algebraic level, the dual $SL(2,C)$ would then be a perfectly well defined symmetry of the BFKL equation. Also, as discussed at the end of the next section, at least the lowest order correction \eqref{eq:ward2d} will not change the commutation of the original and dual $SL(2,C)$, keeping open the possibility that the algebra obtained at leading order describes an exact symmetry of BFKL.

\section{Commuting the original and dual $SL(2,C)$ symmetries}

\no
In $\mathcal{N}=4$ SYM, when taken together, the ordinary and dual superconformal symmetries generate an infinite dimensional Yangian algebra \cite{plefka}. The same Yangian appears in the study of the spectrum of the dilatation operator \cite{dilatationYangian,DolanWittenNappi}, which is believed to define the Hamiltonian of an integrable system \cite{integrabilityN4}, and could therefore explain the great amount of structure observed also for scattering amplitudes. Indeed, it has recently been shown \cite{DrummondKorchemskyGrassmann} that the Grassmannian integral \cite{Grassmannintegral} conjectured to give the amplitudes' leading singularities to all orders is the most general invariant of the Yangian algebra of this kind. Continuing this line of thought, perhaps the original and dual $SL(2,C)$ algebras of BFKL may explain its integrability when taken together. In this section we will check what form the closure of the two $SL(2,C)$ algebras takes.

A Yangian \cite{Yangian,DolanWittenNappi} is generated by a set of elements $J_a^{(0)}$ and $J_{a}^{(1)}$, which we can call level 0 and level 1 respectively, that satisfy
\be
[J_{a}^{(0)},\, J_b^{(0)}]= f_{ab}^{\phantom{ab}c}J_{c}^{(0)} \ , \label{eq:Yangian1}
\ee
and
\be
[J_{a}^{(1)},\, J_b^{(0)}]= f_{ab}^{\phantom{ab}c}J_{c}^{(1)} \ , \label{eq:Yangian}
\ee
for some structure constants $f_{ab}^{\phantom{ab}c}$, as well as the Serre relations
\begin{align}
& [J_{a}^{(1)},\, [J_{b}^{(1)},\, J_{c}^{(0)}]] + [J_{b}^{(1)},\, [J_{c}^{(1)},\, J_{a}^{(0)}]] + [J_{c}^{(1)},\, [J_{a}^{(1)},\, J_{b}^{(0)}]] = \nonumber \\
& \phantom{[J_{a}^{(1)},\, [} h f_{ak}^{\phantom{ak}d}f_{bl}^{\phantom{bl}e}f_{cm}^{\phantom{cm}f}f^{klm}\{ J_{d}^{(0)},\, J_{e}^{(0)},\, J_{f}^{(0)} \}  \ , \label{eq:Serre1}  \\
& [[J_{a}^{(1)},\, J_{b}^{(1)}],\, [J_{c}^{(0)},\, J_{d}^{(1)}]] + [[J_{c}^{(1)},\, J_{d}^{(1)}],\, [J_{a}^{(0)},\, J_{b}^{(1)}]] =  \nonumber \\
& \phantom{[J_{a}^{(1)},\, [}h \left( f_{al}^{\phantom{al}g}f_{bm}^{\phantom{bm}e}f_{kn}^{\phantom{kn}f}f^{lmn}f_{cd}^{\phantom{cd}k} + f_{cl}^{\phantom{cl}g}f_{dm}^{\phantom{dm}e}f_{kn}^{\phantom{kn}f}f^{lmn}f_{ab}^{\phantom{ab}k}  \right) \{ J_{g}^{(0)},\, J_{e}^{(0)},\, J_{f}^{(0)} \}   \, \label{eq:Serre2}
\end{align}
where $h$ is a number which depends on conventions, and where $ \{ \cdot,\, \cdot ,\, \cdot \}  $ is the symmetrized triple product. Satisfying \eqref{eq:Yangian1}-\eqref{eq:Serre2} guarantees that an infinite dimensional algebra is obtained.

In the case of $\mathcal{N}=4$ (where a supersymmetric version of the Serre relations apply), the ordinary $PSU(2,2|4)$ can be taken to make up the level 0 generators, while the level 1 generators can be extracted from the dual superconformal algebra. In \cite{plefka} it was shown that in the representation acting on the scattering amplitudes, where the level 0 generators are written as a sum over external particles as
\be
J_a^{(0)} = \sum_{i}^n J_{ia}^{(0)} \ ,
\ee
the level 1 generators could be written in the bi-local form
\be
J_a^{(1)}=f_{a}^{\phantom{a}bc}\sum_{1\leq j < i \leq n}J_{jb}^{(0)}J_{ic}^{(0)} \ , \label{eq:Ja1}
\ee
where the indices of the structure constants have been raised by the metric of the algebra. This bi-local form guarantees the commutation relation \eqref{eq:Yangian}, and together with some additional requirements satisfied by the representation in question the Serre relations are also implied \cite{DolanWittenNappi}. Recently, it has also been shown that the role of the ordinary and dual algebras can be interchanged, taking the dual superconformal algebra to make up the level 0 generators \cite{DrummondnewYangian}.

\subsection{The original symmetry}

Let us first review the action of the original $SL(2,C)$ symmetry of the BFKL Hamiltonian, discovered by Lipatov in \cite{Lipatov1986}, which we will take to play the role of the level 0 generators $J_a^{(0)}$. It is uncovered by performing a Fourier transform of $f$ into impact parameter space: 
\be
F(\bo{\rho }) = \int d^2 \bo{k}_Ad^2\bo{k}_Bd^2\bo{q}e^{i\left( \bo{\rho}_1 \cdot \bo{k}_A + \bo{\rho}_2 \cdot (\bo{q}-\bo{k}_A)-\bo{\rho}_3 \cdot (\bo{q}-\bo{k}_B) -\bo{\rho}_4 \cdot \bo{k}_B\right)}\frac{f(\bo{k}_A,\, \bo{k}_B,\, \bo{q})}{\bo{k}_B^2(\bo{k}_A-\bo{q})^2} \  , \label{eq:Fourierf}
\ee
The additional factors $\bo{k}_B^2$ and $(\bo{k}_A-\bo{q})^2$, appearing in \eqref{eq:Fourierf} correspond to propagators removed from $f$ in the normalization of the BFKL equation we use. This can be rewritten in the form of an ordinary Fourier transform in terms of incoming momenta $p_i$ as
\be
F(\bo{\rho }) =  \int \prod _i d^2 \bo{p}_i e^{i  \bo{\rho}_i \cdot \bo{p}_i } F(\bo{p}) \  ,
\ee
where
\be
F(\bo{p})=\frac{f(\bo{p})}{\bo{p}_4^2 \bo{p}_2^2} \delta ^{(2)} \left( \sum  \bo{p}_i \right)  \  . \label{eq:F}
\ee

In terms of the complex coordinate $\rho = \rho ^x + i \rho ^y$ one finds an invariance of $F(\rho)$ when
\be
\rho \rightarrow \frac{a + b \rho}{c + d \rho} \ , \label{eq:SL2CL}
\ee
where $a,\, b,\, c,$ and $d$ are complex parameters satisfying $ad-bc=1$. The transformations \eqref{eq:SL2CL} thus represent the group $SL(2,C)$. This group is generated by the transformations
\be
\rho \rightarrow a + \rho \ ,
\ee
corresponding to translations, 
\be
\rho \rightarrow b\rho \ ,
\ee
corresponding to dilatations (when $b$ is real) and rotations (when $b$ is a phase),
\be
\rho \rightarrow \frac{1}{\rho} \ ,
\ee
which is a complex inversion. All of these transformations are standard two-dimensional conformal transformations. With the exception of the complex inversion they all have analogous transformations in the four-dimensional conformal group, acting on a four-vector $x^\mu$. The reason that the inversion is different is that in four-dimensions it appears as
\be
x^\mu \rightarrow \frac{x^\mu}{x^2} \ 
\ee
which would correspond in the two-dimensional case to
\be
\boldsymbol{\rho} \rightarrow \frac{\boldsymbol{\rho}}{\boldsymbol{\rho}^2} \ ,
\ee
which is
\be
\rho \rightarrow \frac{1}{\rho ^\ast} \ ,
\ee
written in complex notation. It is this last transformation that we will refer to as the two-dimensional inversion. It is also a symmetry of BFKL since the kernel is real, implying invariance under $\rho \rightarrow \rho ^\ast$. 

In order to better see the relation with the four-dimensional conformal group, where we can view the $SL(2,C)$ symmetry as the subgroup comprised of dilatations, and rotations, translations and special conformal transformations with indices taking values in the transverse plane, we will now drop the convenient complex notation and instead use two-dimensional vector notation. When we write $\rho ^\mu_i$, the superscript is a two-dimensional index taking the values $x$ and $y$, while the subscript labels the particle, taking values $i=1,\,\ldots,\, n$. In the case of BFKL, $n$ is 4, while it may be higher when we study its extensions.  We will now construct the infinitesimal generators corresponding to the translations, rotations, dilatations and special conformal transformations, following the convention that an infinitesimal generator $J_a$ produces a finite transformation through  $e^{i\xi  J_{a}}$. 

In impact parameter space the generators are given by the standard expressions. Infinitesimal two-dimensional translations, result in the usual expression
\be
P_{\mu} = -i \sum _i \frac{\partial}{\partial \rho_i^\mu} \  \label{eq:Plipatov}
\ee
for the momentum operator, while an infinitesimal dilatation induced the change gives the generator for dilatations as
\be
D = -i \sum _i \rho_i^\mu \frac{\partial}{\partial \rho_i^\mu} \ .
\ee
An infinitesimal (counterclockwise) rotation gives the generator of rotations as
\be
R = -i \epsilon _{\mu \nu}\sum _i \rho_i^\mu \frac{\partial}{\partial \rho_{i\, \nu}} \ ,
\ee
with $\epsilon_{12}=1$.
The special conformal transformations are defined by the usual ITI (Inversion-Translation-Inversion) transformation
giving
\be
K_\mu = -i \sum_i \left(\boldsymbol{\rho}_i^2 \frac{\partial}{\partial \rho_i^\mu} - 2\rho_\mu \rho^\nu \frac{\partial}{\partial \rho_i^\nu} \right) \ . \label{eq:Klipatov}
\ee


Before ending this section we will Fourier transform the generators. The reason is that the dual $SL(2,C)$ is more naturally expressed in the momentum representation, and in order to be able to combine the two symmetries they must be expressed in a common language. The way we perform the transformation is simply to act with the generators on
\be
F(\bo{\rho}) = \int \prod _i\frac{d\boldsymbol{p}_i}{(2\pi)^2}e^{i\bo{\rho}_i \cdot \bo{p}_i}F(\bo{p}) \label{eq:Fourier}
\ee
and rewrite, by partial integration, this as an action on $F(\bo{p})$. For simplicity we will denote the Fourier transformed generators by the same symbol as before.

We obtain
\be
D = i \sum _i \left(2 + p_{i\mu} \frac{\partial}{\partial p_{i\mu}}\right) \ , \label{eq:Dlipatovp}
\ee
were the constant 2 stems from being in two dimensions,
\be
R = -i \epsilon _{\mu \nu}\sum _i p_i^\mu \frac{\partial}{\partial p_{i\, \nu}} \ , \label{eq:Rlipatovp}
\ee
and
\be
K_\mu = \sum_i \left( 4 \frac{\partial}{\partial p_{i}^\mu} + 2 p_i^\nu \frac{\partial}{\partial p_{i}^\nu}\frac{\partial}{\partial p_{i}^\mu} - p_{i\mu}\frac{\partial}{\partial p_{i}^\nu}\frac{\partial}{\partial p_{i\nu}}\right) \ . \label{eq:Klipatovp}
\ee
Here the coefficient in front of $\frac{\partial}{\partial p_{i}^\mu}$ is twice the number of dimensions, and is therefore 4 in our case. And finally, the momentum will of course be
\be
P_{\mu} = \sum_i p_{i\, \mu} \ . \label{eq:Plipatovp}
\ee

\subsection{The dual symmetry in the bi-local form}

\no
We will now show how a piece can be extracted from the dual $SL(2,C)$ symmetry that is (almost) of the form \eqref{eq:Ja1}. As explained above, starting from the set of incoming momenta $\{ \bo{p}_i \}$, $i=1,\, \ldots,\, n$, the the dual $SL(2,C)$ symmetry is uncovered by performing the change of variables
\be
\bo{x}_i -\bo{ x}_{i+1}=\bo{p}_i \ . \label{eq:xp}
\ee
Momentum conservation is automatically satisfied by identifying $\bo{x}_1$ with $\bo{x}_{n+1}$, but since we are interested in the algebraic structure that results from commuting the ordinary and dual symmetries, we will let $\bo{x}_1$ and $\bo{x}_{n+1}$ be independent variables, and include a factor $\delta (\bo{x}_1-\bo{x}_{n+1})$ in the Green's function to impose momentum conservation.

In terms of the $\bo{x}$-variables, the generators of the dual $SL(2,C)$ take the same form as the generators \eqref{eq:Plipatov}-\eqref{eq:Klipatov} do in terms of the $\rho$-variables. It should be noted, though, that just as for the $\mathcal{N}=4$ scattering amplitudes, not all the dual generators are invariances. The translations and the rotations are so, while the dilatations and the special conformal transformations are covariances. One can easily make them them into invariances, however, by shifting them by a constant term. This must be done before they can be combined with the generators \eqref{eq:Plipatov}-\eqref{eq:Klipatov}. First, let us rewrite the generators of the dual algebra in terms of momenta $p$.

The inverse of the change of variables \eqref{eq:xp} is
\be
\bo{x}_i =\bo{ x}_1 - \sum_{j=1}^{i-1}\bo{p}_j \ ,
\ee
where we have kept $\bo{x}_1$ as an independent variable, together with the momenta. Requiring that the $\bo{x}_i$ all be independent, in the sense that
\be
\frac{\partial x^\mu_i}{\partial x^\nu_j}=\delta ^i_j \delta ^\mu_\nu \ 
\ee
then implies that the derivatives with respect to the $x$ should be replaced, when going to the $p$ variables, as
\begin{align}
\frac{\partial}{\partial x^\mu_1} &\rightarrow \frac{\partial}{\partial p^\mu_1} + \frac{\partial}{\partial x^\mu_1} \\
\frac{\partial}{\partial x^\mu_i} &\rightarrow \frac{\partial}{\partial p^\mu_i} - \frac{\partial}{\partial p^\mu_{i-1}} \ , \ i=2,\, \ldots ,\, n \\
\frac{\partial}{\partial x^\mu_{n+1}} &\rightarrow -\frac{\partial}{\partial p^\mu_n} \ .
\end{align}
Performing these substitutions, the dilatation operator becomes
\begin{align}
D^{(D)} &= -i \sum _{i=1}^{n+1}\bo{x}_i \cdot \frac{\partial}{\partial \bo{x}_i} = \nonumber \\
&= -i\left( \bo{x}_1 \cdot \frac{\partial}{\partial \bo{x}_1}+ \bo{x}_1 \cdot \frac{\partial}{\partial \bo{p}_1} + (\bo{x}_1 - \bo{p}_1) \cdot \left(\frac{\partial}{\partial \bo{p}_2} - \frac{\partial}{\partial \bo{p}_1} \right)+ \cdots  \right. \nonumber \\
& \;\;\;\;\;\;\;\;\;\;\;\; \cdots \left. - (\bo{x}_1 - \bo{p}_1 - \cdots -\bo{p}_n) \cdot \frac{\partial}{\partial \bo{p}_n} \right) = \nonumber \\
&= -i\left( \bo{x}_1 \cdot \frac{\partial}{\partial \bo{x}_1} + \sum_{i=1}^n \bo{p}_i \cdot \frac{\partial}{\partial \bo{p}_i} \right) \ ,
\end{align}
where we have introduced the superscript $(D)$ to distinguish this set of generators from that of the ordinary conformal symmetry. We see that when acting on an object independent of $\bo{x}_1$ (or,  equivalently, which only depends on the momenta), this dilatation operator is the same, up to a change of sign and a shift by a constant, as the original dilatation operator \eqref{eq:Dlipatovp}. 

We do not obtain anything new from translations and rotations, either. Up to terms containing derivatives with respect to $x_1$ the former are represented by the identity, while the latter give the same generator as the original symmetry \eqref{eq:Rlipatovp}. The only new symmetry comes from the special conformal generators, mimicking the structure appearing in the scattering amplitudes of $\mathcal{N}=4$ SYM, in the form of the original and dual conformal generators.

The special conformal generators take the form
\begin{align}
iK^{(D)}_\mu &= \bo{x}_1^2 \frac{\partial}{\partial x_1^\mu} - 2x_{1\mu} \left( \bo{x}_1\cdot \frac{\partial}{\partial \bo{x}_1} + \sum_i \bo{p}_i\cdot \frac{\partial}{\partial \bo{p}_i} \right)- 2\sum_ix_1^\nu \left(p_{i\mu}\frac{\partial}{\partial p^\nu_i} - p_{i\nu}\frac{\partial}{\partial p^\mu_i}  \right) + \nonumber \\
&+\sum_i\left(2p_{i\mu}\bo{p}_i\cdot \frac{\partial}{\partial \bo{p}_i}  - \bo{p}_i^2 \frac{\partial}{\partial p^\mu_i}\right) + 2 \sum_{i=2}^n \sum_{j=1}^{i-1}\left( p_{i\mu}\bo{p}_j \cdot \frac{\partial}{\partial \bo{p}_i} +  p_{j\mu} \bo{p}_i \cdot \frac{\partial}{\partial \bo{p}_i}  - \bo{p}_i \cdot \bo{p}_j \frac{\partial}{\partial p^\mu_i} \right) \ . \label{eq:K}
\end{align}

Again, when acting on a physical object, which only depends on the momenta, all of the derivatives with respect to $x_1$ can be dropped. Furthermore, the third term can also be dropped due to the rotational invariance. The remaining $x_1$-dependence is given by
\be
- 2x_{1\mu}\sum_i \bo{p}_i\cdot \frac{\partial}{\partial \bo{p}_i} \ . \label{eq:partD}
\ee

As mentioned above,  $K^{(D)}_\mu$ is not an invariance of the gluon Green's function, and must be corrected by a constant piece. In order to do so we will hereafter restrict our attention to the case $n=4$ of BFKL for which we know the transformation properties. Under dual inversions $f \rightarrow \bo{x}_2^2\bo{x}_4^2 f$, and from \eqref{eq:F} we get 
\be
F(\bo{p}) = \frac{f(\bo{x})}{\bo{x}_{14}^2\bo{x}_{23}^2}\delta ^{(2)}(\bo{x}_{15}) \rightarrow \bo{x}_1^6\bo{x}_2^4\bo{x}_3^2\bo{x}_4^4 F(\bo{p}) \ . \label{eq:Finv}
\ee
This implies that acting on $F(\bo{p})$ with $iK^{(D)}_\mu$ produces the factor
\be
6 x_{1\mu}+4x_{2\mu}+2x_{3\mu}+4x_{4\mu}= 16 x_{1\mu} -10p_{1\mu} - 6 p_{2\mu} - 4p_{3\mu} \ ,
\ee
and we see that the gluon Green's function will be invariant under the combination 
\be
iK^{(D)}_\mu  +  10p_{1\mu} + 6 p_{2\mu} + 4p_{3\mu} - 16x_{1\mu} \ .
\ee
Here, we see that the $-16x_{1\mu}$ term is precisely what is needed in order to eliminate the remaining $\bo{x}_1$ dependence completely. When added to \eqref{eq:partD}, one obtains a term which is proportional to \eqref{eq:Dlipatovp} and can be dropped. In the end, we extract the new generator
\begin{align}
\hat{K}_\mu &= \sum_i\left(2p_{i\mu}\bo{p}_i\cdot \frac{\partial}{\partial \bo{p}_i}  - \bo{p}_i^2 \frac{\partial}{\partial p^\mu_i}\right) +  10p_{1\mu} + 6 p_{2\mu} + 4p_{3\mu} + \nonumber \\
&+ 2 \sum_{i=2}^n \sum_{j=1}^{i-1}\left( p_{i\mu}\bo{p}_j \cdot \frac{\partial}{\partial \bo{p}_i} +  p_{j\mu} \bo{p}_i \cdot \frac{\partial}{\partial \bo{p}_i}  - \bo{p}_i \cdot \bo{p}_j \frac{\partial}{\partial p^\mu_i} \right) \  \label{eq:Khat}
\end{align}
from the dual $SL(2,C)$.

We will now go on to remove a part of \eqref{eq:Khat}, which itself annihilates the Green's function, so that what is left is almost of the bi-local form \eqref{eq:Ja1}. The relevant bi-local operators turn out to be those where the index $a$ corresponds to the two-dimensional momentum $P_\mu$. Using the metric $g_{ab}=\frac{1}{4}f_{ac}^{\phantom{ac}e}f_{be}^{\phantom{be}c}$ for the algebra to raise indices, where the constant $1/4$ is chosen for convenience, and inserting the level 0 generators \eqref{eq:Dlipatovp}-\eqref{eq:Plipatovp} into \eqref{eq:Ja1} we have the operator
\begin{align} 
\tilde{J}_\mu^{(1)} &= -i \sum_{1\leq j < i \leq n}\left[  P_{j\mu}D_i - \epsilon _{\mu \nu} P_{j \nu} R_i   -  (i \leftrightarrow j) \right]   = \nonumber \\
&= \sum_{1\leq j < i \leq n}\left[  p_{j\mu}\left( 2 + p_i^\rho \PD{p_i^\rho} \right) + \epsilon _{\mu \nu} p_{j}^\nu\epsilon_{\rho \lambda}p_i^\rho \PD{p_{i\lambda}}   -  (i \leftrightarrow j) \right]  \label{eq:J13} \ ,
\end{align}
where a tilde is added to indicate that we do not know yet if these are symmetries of the Green's function.

In order to arrive at \eqref{eq:J13} let us start by splitting the last sum in \eqref{eq:Khat} into two equal pieces. One of the pieces we leave in it's current form, while using $P_\mu=\sum p_{i\mu}$ we rewrite the second piece as
\begin{align}
\sum_{i=2}^n&\left( p_{i\mu}(\bo{P}-\bo{p}_{i} - \cdots -\bo{p}_{n})\cdot \frac{\partial}{\partial \bo{p}_i} +  (P_\mu -p_{i\mu} - \cdots -p_{n\mu})\bo{p}_i \cdot \frac{\partial}{\partial \bo{p}_i} -\right. \nonumber \\
&\left. - \bo{p}_i \cdot (\bo{P}-\bo{p}_i - \cdots - \bo{p}_{n})\frac{\partial}{\partial p^\mu_i} \right) \ . \label{eq:Khatpiece}
\end{align}
The nested sums run from index $i$ to $n$, and we can cancel the terms corresponding to index $i$ with terms from the first line of \eqref{eq:Khat}. The only part that then remains from the sum in the first line is the term corresponding to $i=1$,  
\begin{align}
& 2p_{1\mu}\bo{p}_1\cdot \PD{\bo{p}_1}-\bo{p}_1^2\PD{p_1^\mu}= \nonumber \\
& p_{1\mu}(\bo{P}-\bo{p}_{2} - \cdots -\bo{p}_{n})\cdot \frac{\partial}{\partial\bo{p}_1} +(P_\mu -p_{2\mu} - \cdots -p_{n\mu})\bo{p}_1 \cdot \frac{\partial}{\partial \bo{p}_1} - \nonumber \\
& - \bo{p}_1 \cdot (\bo{P}-\bo{p}_2 - \cdots - \bo{p}_{n})\frac{\partial}{\partial p^\mu_1}  \ .
\end{align}
These terms complete the sum in \eqref{eq:Khatpiece} so that it starts from $i=1$. Altogether, we find
\begin{align}
\hat{K}_\mu &= \sum_{i=1}^n\left(p_{i\mu}\bo{P}\cdot \frac{\partial}{\partial \bo{p}_i}+P_{\mu}\bo{p}_i \cdot \PD{\bo{p}_i}  - \bo{p}_i \cdot \bo{P} \frac{\partial}{\partial p^\mu_i}\right)  +  10p_{1\mu} + 6 p_{2\mu} + 4p_{3\mu} + \nonumber \\
&+ \sum_{i=2}^n \sum_{j=1}^{i-1}\left( p_{i\mu}\bo{p}_j \cdot \frac{\partial}{\partial \bo{p}_i} +  p_{j\mu} \bo{p}_i \cdot \frac{\partial}{\partial \bo{p}_i}  - \bo{p}_i \cdot \bo{p}_j \frac{\partial}{\partial p^\mu_i} \right)  - \nonumber \\
&-\sum_{i=1}^{n-1} \sum_{j=i+1}^{n}\left( p_{i\mu}\bo{p}_j \cdot \frac{\partial}{\partial \bo{p}_i} +  p_{j\mu} \bo{p}_i \cdot \frac{\partial}{\partial \bo{p}_i}  - \bo{p}_i \cdot \bo{p}_j \frac{\partial}{\partial p^\mu_i} \right) \ . 
\label{eq:Khat3}
\end{align}
The first sum will itself annihilate the Green's function. In fact, it will annihilate any function of the form $\delta ^{(2)}(\bo{P}) \, f(\{ \bo{p}_i \})$, which includes a momentum conserving delta function. The factors of $P$ make all terms in which the derivatives act on $f$ vanish so only the terms in which the derivatives act on the delta function remain. For the same reason, when the derivatives are moved off the delta function by partial integration they must act on the factors of $P$ in order to get a non-vanishing contribution. We are then left with
\be -\sum_{i=1}^n\left(p_{i\mu}2 +p_{i\mu}  - p_{i\mu}  \right)\delta ^{(2)}(\bo{P}) \, f(\{ \bo{p}_i \}) \ , \ee
which once again vanishes due to the delta function.

Extracting the first term from \eqref{eq:Khat3},  what remains is
\be
\sum_{1\leq j < i \leq n} \left( p_{i\mu}\bo{p}_j \cdot \PD{\bo{p}_i} + p_{j\mu}\bo{p}_i \cdot \PD{\bo{p}_i} - \bo{p}_i \cdot \bo{p}_j \PD{p_{i}^\mu}  - (i \leftrightarrow j)\right) +  10p_{1\mu} + 6 p_{2\mu} + 4p_{3\mu}  \ ,
\ee
which can be rewritten (in two dimensions) as
\be
\sum_{1\leq j < i \leq n} \left(  p_{j\mu}p^\rho_i \PD{p^\rho_i}  + \epsilon_{\mu \nu}p_{j}^\nu \epsilon_{\rho \sigma}p_i^\rho \PD{p_{i\sigma}}  - (i \leftrightarrow j)\right) +  10p_{1\mu} + 6 p_{2\mu} + 4p_{3\mu} \ , \label{eq:Khat4}
\ee
reproducing a large portion of the terms in \eqref{eq:J13}. The only one that remains unaccounted for is 
\be
 \sum_{1\leq j < i \leq n} 2  (p_{j\mu} - p_{i\mu})= 4 \left( (n-1)p_{1\mu} + (n-2)p_{2\mu} + \cdots + p_{n-1} - (n-1)P_\mu \right) \ .
\ee
For the case that interests us, which is $n=4$, this becomes, after discarding the term proportional to $P_\mu$,
\be
12 p_{1\mu} + 8 p_{2\mu} + 4p_{3\mu}  \ ,
\ee
which is almost, but not exactly, equal to the term $10p_{1\mu} + 6 p_{2\mu} + 4p_{3\mu}$, present in \eqref{eq:Khat4}. The difference is $2p_{1\mu}+2p_{2\mu} \equiv 2q_\mu$. 

We must thus conclude that, in contrast to the case of $\mathcal{N}=4$ SYM, the bi-local operators $\tilde{J}_\mu^{(1)}$ are in general not symmetries of the Green's function. We instead have the symmetries
\be
J^{(1)}_{\mu}\equiv \tilde{J}_{\mu}^{(1)} - 2 q_\mu \ . \label{eq:P1}
\ee
Commuting these generators with the level 0 algebra gives the rest of the "level 1" generators, all of which can be written as as
\be
J^{(1)}_a \equiv \tilde{J}^{(1)}_a - 2J^{(0)}_{(12)a} \ , \label{eq:Ja1sym}
\ee
where
\be
J^{(0)}_{(12)a}\equiv \sum_{i=1}^2J^{(0)}_{ia} \ 
\ee
are the level 0 generators restricted to the first two (upper) momenta, and where the $\tilde{J}^{(1)}_a$ are defined by the bi-local formula.

It should be noted, that the reason that we are forced to deform the $J_a^{(1)}$ generators is the non-symmetrical action \eqref{eq:Finv} of the dual conformal inversions. The $\tilde{J}_{a}^{(1)}$ would have been symmetries if $F(p)$ had produced a factor $x_1^4x_2^4x_3^4x_4^4$ under inversions \footnote{This should be compared with the amplitudes of $\mathcal{N}=4$ for which the inversion produces the factor $x_1^2\cdots x_n^2$, consistent with the bi-local formula because the relevant constants appearing in the generators (such as the constant in the dilatation operator) are half of what they are in the present case.}. 

Notable is that in the forward limit $\bo{q}=0$, and then the bi-local operators will indeed be symmetries. This case is particularly important, since it is related, via the optical theorem, to the total cross section. In \cite{SeverVieira} it was discussed how IR safe quantities, which includes the total cross section, retain the superconformal symmetry of $\mathcal{N}=4$. By contrast, in order to promote the superconformal symmetry of tree amplitudes beyond tree level, an analysis carried out in \cite{BeisertoneloopYangian}, the generators must be non-trivially deformed. This suggests that a more direct link to the integrable structure of $\mathcal{N}=4$ might be found by looking at the forward limit.

\subsection{Is there a Yangian?}

So, does the original $SL(2,C)$ together with \eqref{eq:Ja1sym} generate a Yangian? It follows from the form of the bi-local formula that the Yangian commutation relations \eqref{eq:Yangian} are satisfied with the $\tilde{J}^{(1)}$, implying that they are also obeyed by the full $J^{(1)}$, since the additional $2J^{(0)}_{(12)a} $ piece simply gives the level 0 algebra restricted to particles 1 and 2. The key issue is therefore whether the Serre relations \eqref{eq:Serre1}-\eqref{eq:Serre2} are satisfied. 

For many algebras, such as $PSU(2,2|4)$ the first Serre relation implies the second one \cite{DolanWittenNappi}.   And in our case the first Serre relation is indeed satisfied. However, for $SL(2,C)$ it is trivially satisfied (its structure constants imply it take the form $0=0$), meaning that it is the second Serre relation that must be checked. Unfortunately, it turns out to not be satisfied. From the form of \eqref{eq:Serre1} and \eqref{eq:Serre2} it is clear that if the right hand side of the first relation is zero, the right hand side of the second relation must also be so. But, a quick check reveals that the left hand side becomes non-zero when using the generators \eqref{eq:Ja1sym}. So it does not seem that the closure of the original and dual $SL(2,C)$-algebras give a Yangian algebra \footnote{It might still be possible to realize a Yangian using some other prescription for the level 1 algebra than that given by the bi-local form, or \eqref{eq:Ja1sym}, although it does not seem likely.}. This does not mean that the algebra cannot be infinite dimensional, though. The generators $\left\{ J_a^{(0)},\, J_a^{(1)} \right\}$ do not close under commutation, and from their form it seems unlikely that their closure would be finite-dimensional. 

An important issue is also whether the algebra itself is affected when the representation of the dual $SL(2,C)$ is deformed to $J_a^{(1)}(\hat{\alpha})$ in order to take into consideration the anomaly \eqref{eq:ward2d}. The full form of the deformed generator becomes
\be
J^{(1)}_a (\hat{\alpha}) = \tilde{J}^{(1)}_a - 2J^{(0)}_{(12)a} - 2 \hat{\alpha}J^{(0)}_{(13)a} + \mathcal{O}\left( \hat{\alpha}^2 \right) \ , \label{eq:Ja1symdef}
\ee
where $J^{(0)}_{(13)a} $ is defined analogously to $J^{(0)}_{(12)a} $. Interestingly, in the same way that the shift in \eqref{eq:Ja1sym} by $-2J^{(0)}_{(12)a}$ does not alter the commutation relation \eqref{eq:Yangian}, neither does the change in \eqref{eq:Ja1symdef}. Also when considering the rest of the algebra, generated by $J_a^{(0)}$ and $J_a^{(1)}(\hat{\alpha})$ it does not seem that the commutation relations are altered by the introduction of the coupling dependence, meaning that the algebra obtained at lowest order may very well be exact to all orders of BFKL. We will, however, leave the exact determination of this algebraic structure to future studies.

\section{Discussion}

\no
In this article we have continued the study of how the symmetries of $\mathcal{N}=4$ SYM may explain the integrability found in the Regge limit of QCD, examining the BFKL equation. The conclusion is that even though much of the structure present in the four-dimensional gauge theory remains (original and dual $SL(2,C)$ as remnants of ordinary and dual conformal symmetry, generators in the bi-local form, etc), it seems that, algebraically, things get considerably more complicated in this limit. In particular, the Yangian symmetry is broken, and the dual $SL(2,C)$ does not satisfy a simple all-loop anomalous Ward identity, although the possibility remains that the algebra itself be coupling independent to all orders. These difficulties probably result from breaking the symmetries corresponding to the longitudinal and fermionic generators upon taking the Regge limit, since the structure of the Yangian of $\mathcal{N}=4$ seems intimately related to supersymmetry (the bi-local form of the level 1 generators is, for example, only compatible with the cyclicity of the amplitudes for a set of superalgebras \cite{plefka}.).

There are several possible directions for future research. It would, of course, be nice to find an all-order representation of the dual $SL(2,C)$. But perhaps a more immediate issue is to extend this analysis to the full generalized LLA spin chain of \cite{GLLAspinchain,FaddeevKorchemsky}, and see if the structure of conserved charges can be related to the closure of the two $SL(2,C)$ algebras. Another possibility is to study the BK equation \cite{BK}, giving unitarity corrections to BFKL by taking into account transitions from 2 to 4 reggeized gluons. Since it was shown in \cite{Nosotros} that the $2$ to $4$ reggeized gluon vertex has the same dual $SL(2,C)$ covariance as the $2$ to $2$ vertex present in BFKL, and in \cite{reggevertex2a4} that it has the original $SL(2,C)$-invariance, it seems likely that the entire equation has this symmetry.

A more fundamental step that remains in order to understand the symmetries of the Regge limit is precisely to elucidate exactly how, not only the dual, but also the original $SL(2,C)$ symmetry of Lipatov is implied by the symmetries of the four dimensional gauge theory. The solution to this problem is bound to be interesting, especially at NLLA, since the conformal symmetry is broken at loop level in $\mathcal{N}=4$. One source of the breaking is the introduction of an infrared regulator. Since BFKL is IR finite one could think that this is not a problem, but we have seen in the form of the breaking of the dual $SL(2,C)$ that IR effects can still cause problems. A second, trickier problem is that the generators of the superconformal symmetry must be deformed in a non-trivial way at one-loop and beyond \cite{BeisertoneloopYangian}. This deformation must somehow disappear before arriving at the $SL(2,C)$ symmetry of the Regge limit, since the latter is exact to all orders, also at NLLA, with no deformation present.

\vspace{10mm}
\centerline{\bf Acknowledgments}

We are grateful to J. Drummond, C. G\'omez and A. Sabio Vera for comments and discussions. The work of J. G. is supported by a Spanish FPU grant.

\appendix

\section{Using dual $SL(2,C)$ to evaluate the next-to-leading order integral.}

\label{app:nfintegral}

\no
In this appendix we will show how the form of the integral 
\be
I_1\equiv \frac{1}{\pi}\int \frac{d^2\bo{x}_I \, \bo{x}_{1I}^2}{\bo{x}_{2I}^2 \bo{x}_{3I}^2 \bo{x}_{4I}^2} \  \label{eq:nfintegral}
\ee
can be restricted by using dual $SL(2,C)$ symmetry, thus exemplifying how this symmetry can play a role in simplifying calculations. First, we note that the integrand of $I_1$ has the same behavior close to the singularities at $\bo{x}_2$, $\bo{x}_3$ and $\bo{x}_4$, and the same transformation properties under dual $SL(2,C)$ transformations as the integrand of
\be
I_2 \equiv \frac{1}{2\pi}\frac{\bo{x}_{12}^2}{\bo{x}_{23}^2 \bo{x}_{24}^2} \int d^2\bo{x}_I \left[ (1+v-u)\frac{\bo{x}_{23}^2}{\bo{x}_{2I}^2 \bo{x}_{3I}^2} + (1-v+u)\frac{\bo{x}_{24}^2}{\bo{x}_{2I}^2 \bo{x}_{4I}^2} + (-1+v+u)\frac{\bo{x}_{34}^2}{\bo{x}_{3I}^2 \bo{x}_{4I}^2}  \right] \ ,
\ee
which can be evaluated directly in dimensional regularization using \eqref{eq:trajectorydimreg}, where
\be
u = \frac{\bo{x}_{14}^2\bo{x}_{23}^2}{\bo{x}_{12}^2\bo{x}_{34}^2} \; \;\;\ \text{and} \;\;\; v = \frac{\bo{x}_{13}^2\bo{x}_{24}^2}{\bo{x}_{12}^2\bo{x}_{34}^2} \ 
\ee
are the two independent dual $SL(2,C)$ invariants that can be formed from $\bo{x}_1,\, \ldots ,\, \bo{x}_4$. This implies that the integral $I_1-I_2$ will be finite and that we can apply the dual conformal symmetry to conclude that
\be
I_1-I_2 = \frac{\bo{x}_{12}^2}{\bo{x}_{23}^2 \bo{x}_{24}^2} I(u,\, v) \ , \label{eq:I1mI2}
\ee
for some function $I(u,\, v)$ of the conformal invariants. Next, we observe that both $I_1$ and $I_2$ are symmetric under the interchanges $2\leftrightarrow 3$, producing
\be
u \leftrightarrow \frac{u}{v} \ , \; v \leftrightarrow \frac{1}{v} \ \text{and} \; \frac{\bo{x}_{12}^2}{\bo{x}_{23}^2 \bo{x}_{24}^2} \leftrightarrow v \frac{\bo{x}_{12}^2}{\bo{x}_{23}^2 \bo{x}_{24}^2}  \ , \label{eq:23swap}
\ee
as well as $3 \leftrightarrow 4$ which gives 
\be
u\leftrightarrow v \ . \label{eq:34swap}
\ee
When applied to \eqref{eq:I1mI2} these symmetries give
\be
I(u,\, v)=v I \left( \frac{u}{v},\, \frac{1}{v} \right) \;\;\;\ \text{and} \;\;\; I(u,\, v)=I(v,\, u) \ , \label{eq:symI}
\ee
which implies that
\be
I(u,\, v) = (1+u+v)G(\xi_1 ,\, \xi_2) \ ,
\ee
for some function $G$, where 
\be
\xi_1 = \frac{u+v+uv}{(1+u+v)^2} \; \;\;\ \text{and} \;\;\;  \xi_2 = \frac{uv}{(1+u+v)^3} \ 
\ee
are two independent invariants of \eqref{eq:23swap} and \eqref{eq:34swap}. 

The difference between the two integrals $I_1$ and $I_2$ is further restricted by noting that for some values of $\bo{x}_1,\, \ldots ,\, \bo{x}_4$, their integrands coincide for all values of the integration variable. This occurs, for example, if $\bo{x}_1$ equals one of the other $\bo{x}_i$, translating to $G(1/4 ,\, 0) = 0$. In fact, it can be shown that the two integrands are equal for all $\bo{x}_I$ precisely when
\be
\frac{u+v+uv}{(1+u+v)^2} = 1/4 \ , \label{eq:integrandequal}
\ee
implying that $G(1/4 ,\, \xi_2) = 0 $, for arbitrary $\xi_2$. As a consequence we can write
\be
I(u,\, v) = (1+u+v)\left(4 \frac{u+v+uv}{(1+u+v)^2}-1\right)H(\xi_1 ,\, \xi_2) = \frac{2(u+v) - (u-v)^2-1}{1+u+v} H(\xi_1 ,\, \xi_2) \ , \label{eq:Ifinal}
\ee
where $H$ must be finite, for all $\xi_2$, when $\xi_1 \rightarrow 1/4$.

It does not seem that we can obtain anything more from symmetry considerations alone. Still, from equations \eqref{eq:I1mI2} and \eqref{eq:Ifinal} we see that the form of $I_1$ is greatly restricted. It should then not come as a surprise that in fact $I_1=I_2$, as can easily be checked for any arbitrary choice of the $\bo{x}_i$. 

This analysis can also be applied to restrict the form of more complicated integrals. For example, at the third iteration of the BFKL equation, integrals such as
\be
I_3 = \frac{1}{\pi}\int \frac{d^2\bo{x}_I \, \bo{x}_{1I}^2}{\bo{x}_{2I}^2 \bo{x}_{3I}^2 \bo{x}_{4I}^2}  \ln \left( \frac{\bo{x}_{4I}^2}{\bo{x}_{1I}^2} \right) \ 
\ee
appear. If we define $I_4$ by exchanging $\frac{1}{\pi}\frac{ \bo{x}_{1I}^2}{\bo{x}_{2I}^2 \bo{x}_{3I}^2 \bo{x}_{4I}^2}$ for the integrand of $I_2$, then $I_4-I_3$ will once again be restricted by dual $SL(2,C)$ symmetry, symmetry under the exchange of $2\leftrightarrow 3$, and the condition that the difference vanishes when \eqref{eq:integrandequal} is satisfied. 



\end{document}